# Survey of image processing settings used for mammography systems in the United Kingdom: how variable is it?

Alistair Mackenzie*[a], John Loveland[a], Ruben van Engen[b]

[a]NCCPM, Medical Physics Department, Royal Surrey NHS Foundation Trust, Guildford, UK, GU2 7XX; [b]Dutch Expert Centre for Screening (LRCB), Wijchenseweg 101, 6538 SW, Nijmegen, The Netherlands

## ABSTRACT

The aim was to undertake a national survey of the setup of mammography imaging systems in the UK, we were particularly interested in image processing and software version. We created a program that can extract selected tags from the DICOM header. 28 medical physics departments used the program on processed images of the TORMAM phantom acquired since 2023 and this produced data for 497 systems. We received data for 7 different models of mammography systems. We found that currently in use each model had between 2 and 7 different versions of software for the acquisition workstation. Each of the systems had multiple versions of image processing settings, a preliminary investigation with TORMAM demonstrated large differences in the appearance of the image for the same X-ray model. The Fujifilm, GE and Siemens systems showed differences in the setup of the dose levels. In addition to these settings there were differences in the paddles used and grid type. Our snapshot of system set up showed that there is a potential for the images to appear differently according to the settings seen in the headers. These differences may affect the outcomes of AI and also human readers. Thus the introduction of AI must take these differences into consideration and the inevitably changes of settings in the future. There are responsibilities on AI suppliers, physics, mammographic equipment manufacturers, and breast-screening units to manage the use of AI and ensure the outcomes of breast screening are not adversely affected by the set-up of equipment.

## 1. INTRODUCTION

There are a number of different ways that Artificial Intelligence (AI) may be used in mammography. AI readers could be used to assist human readers, to triage low risk cases, or to completely replace a human reader. Furthermore AI algorithms have shown promising results for both risk prediction, density estimation and quality control.[1] If AI is successfully implemented then there could be great benefits to breast screening programmes. There is the potential to improve the decision making process and/or require fewer resources.

Virtually all of the artificial intelligence (AI) software in mammography is trained using 'For Presentation' images, i.e. following image processing intended for human readers. There are a range of processing options available including noise reduction, edge enhancement, frequency processing, and look up tables to match the visual response and dynamic range of the eye. AI companies design their product to work with these images as few imaging departments will send the 'For Processing' images to their PACS and so those images will only be stored temporarily on the acquisition workstation. From an information science point of view, although the processing attempts to optimise the appearance of the image for the human viewer, some information is likely to be lost during the image processing, even if it is a small amount and structures might appear differently depending on the applied processing techniques.

This means that the training of AI needs to account for not only the system, but also different versions of image processing. Many of the centres have image processing set up for their requirements within the possibilities offered by the vendor. However, it is likely that not all centres will have their image processing set up optimally and similar to other centres. Some suppliers may even adjust the processing for the individual preference of the screening unit. There is unlikely to be any ideal image processing, but each will have their strengths, weaknesses and appearance. The personal experience of the authors is that some image processing packages/settings could be improved. It has been shown that image processing affects the decision making of human readers, in particular for the task of detecting lesions that should be recalled for further investigation.[2,3]

Readers can adapt to differences in image processing, but can AI products still perform well if image processing is changed? There is some evidence that small changes in image processing, hardly visible to the human eye, impact the effectiveness of AI.[4] AI-software is a blackbox and so it is not known which image features influence the AI decision; potentially these image features could be related to the software and hardware which is used. In addition to image processing there are other differences that could potentially affect the appearance of images and thus AI, such as dose level, changes to the detector design, detector ageing, type of anti-scatter grid or type of compression paddle. These differences could potentially complicate roll out of AI products on a national scale.

We wish to investigate the effect of processing on AI trained for on a mammographic system. The first step is to investigate what differences exist between systems of the same brand in a screening programme at one time. The aim of this work is to perform a snapshot survey of mammography systems in the United Kingdom and ascertain the heterogeneity of the setups of the systems in terms of image processing, image acquisition, and hardware differences (e.g. anti-scatter grids). These parameters may influence the perception of the clinical images and the performance of AI products. This study shows the results for four manufacturers.

## 2. METHODS

Table 1. Data extracted from DICOM header for Fujifilm (F), GE (G), Hologic (H) and Siemens (S)

| Manufacturer | DICOM tags |
| --- | --- |
| Presentation Intent Type | 0008x,0068x |
| Model name | 0008x,1090x |
| Institution name | 0008x,0080x |
| Station name | 0008x,1010x |
| Acquisition date | 0008x,0022x |
| Derivation Description | 0008x,2111x |
| Software Version | 0018x,1020x |
| Image processing | See Table 2 |
| Dose level | 0018x,7062x (G,S), 0019x,1049x (H) |
| Anti-scatter grid | 0019x,109Cx |
| Compression paddle | 0018x,11A4x |

**National survey**

A national survey of image processing used in mammography systems was undertaken. The DICOM conformance statements for mammographic equipment were examined for the four main manufacturers used in the UK. We also examined the DICOM header of each of the systems and identified tags that may be of interest. Ninety-three DICOM tags associated with settings that may affect image presentation and image quality and also tags that identify the system were then selected. The DicomTagExtractor (available on  https://medphys.royalsurrey.nhs.uk/nccpm/?s=breast-dose and used for dose surveys) was adapted to strip these tags from images, if present in the header. The tags that were primarily used in this work are shown in Table 1. A good description of DICOM tags can be found in Riddle and Pickens.[5]

We wrote to each of 35 medical physics departments who undertake quality control testing of mammography systems in the UK breast screening programmes and other mammography imaging departments. Most medical physics departments in the UK will acquire images of the TORMAM phantom (Leeds Test Objects, Leeds, UK) using clinical processing as part of routine quality control. An example of the TORMAM phantom is shown in **Error! Reference source not found.**, half of it has breast like structures with inserted calcification clusters, while the other half contains geometric shapes designed to replicate some of cancer's relevant features. NCCPM asked the medical physics departments to collate 'FOR PRESENTATION' images of the TORMAM already acquired on the mammography systems. We did request some images from a few departments for comparison purposes. The TORMAM has a lot of breast like structures, however, we know that image processing will depend on the object being imaged and so the processing may not function in the same way as for the test object as for a breast. It may give an indication of the differences.

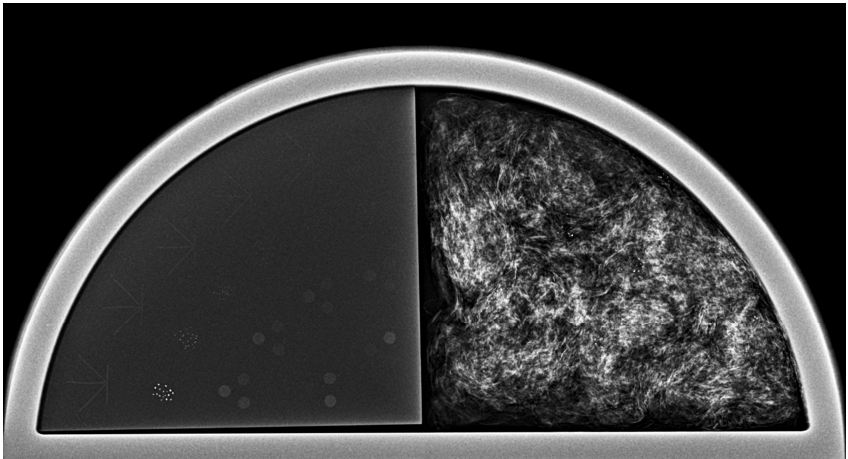

Figure 1. Image of TORMAM.

**Data analysis**

The data was then collated and filtered to only include 'FOR PRESENTATION' images ('Derivation Description' tag), acquired since 1st Jan 2023 ('Acquisition Date' tag), full field digital mammography, only include mammography units (not prone), most recent result for each system. The results of the data in the tags were compared between the imaging systems.

### 3. RESULTS

As part of the survey, 28 out of 35 physics services sent data from at least one system. Not all physics groups use TORMAM or keep a copy of TORMAM images, and so they were unable to take part. In total, we received data for 497 systems surveyed between January 2023 and February 2024. The split by model is shown in Table 2.

**Software versions**

The software version was found in tag 0018x,1020x for the acquisition workstation. Table 2 shows that each system model has between 2 and 7 different software versions across the UK.

Table 2. Number of systems data collected from UK, number of software version used and the number of image processing settings and the tags where the information is held

| Unit | Number | No. software versions | No. of image processing settings | Tag(s) for image processing |
|---|---|---|---|---|
| Fujifilm Amulet | 7 | 7 | 5 | 0008x,2111x |
| GE Essential | 19 | 6 | 3 | 0018,1400x, 0018,7006x |
| GE Pristina | 116 | 4 | 3 | 0018,1400x, 0018,7006x |
| Hologic Selenia Dimensions | 128 | 7 | 4 | 0019,1062x |
| Hologic 3Dimensions | 151 | 4 | 5 | 0019,1062x |
| Siemens Inspiration | 43 | 6 | 10 | 0019,1010x |
| Siemens Revelation | 33 | 2 | 6 | 0019,1010x |

**Image processing**

Table 2 shows the number of image processing settings used for each model. It is noted, that the seven Fujifilm systems have five different image processing setups. These images have a large difference in appearance, in particular in terms of contrast. For the GE medical systems, there are five different processing settings between the Essential and Pristina. There is also an MTF compensation that is applied to GE images. The GE system also has three levels of MTF correction levels (2.1,2.2,2.3) that can be applied to the image. All three were seen in the data both for the Essential and the Pristina.

Siemens have a number of 'Flavors' that can be set up for the Siemens Revelation, in this survey two 'flavors' were in use, but also had a Gen 1 and Gen 2 classification and so had names of F0_Gen1, F0_Gen2, F1_Gen1 and F1_Gen2. For the Inspiration, the image processing settings were called Contrast Levels: Standard_OV2, Med, MedHigh and High. There were also a number of underlying tags describing the image processing which varied with the processing name. Interestingly, there was one centre that appeared to have bespoke image processing (named after that centre) with processing parameters distinct from other systems in the UK.

The Hologic tag (0019x,1097x) is very long with up to 145 separate sub-tags and contain image processing factors and system settings. The number of tags varies with software versions: AWS:1.8, 1.9, 1.10, and 1.11 having 125, 125, 134 and 145 sub-tags respectively. Most of the values of the sub-tags were the same, although some were set up differently. We know that there are some options are for setting the image contrast and appearance of skin line.

Figures 2 and 3 shows examples from two systems when the values in the image processing tags are different. The TORMAM images appear different. It will be of interest to see how the processing changes the appearance of clinical images.

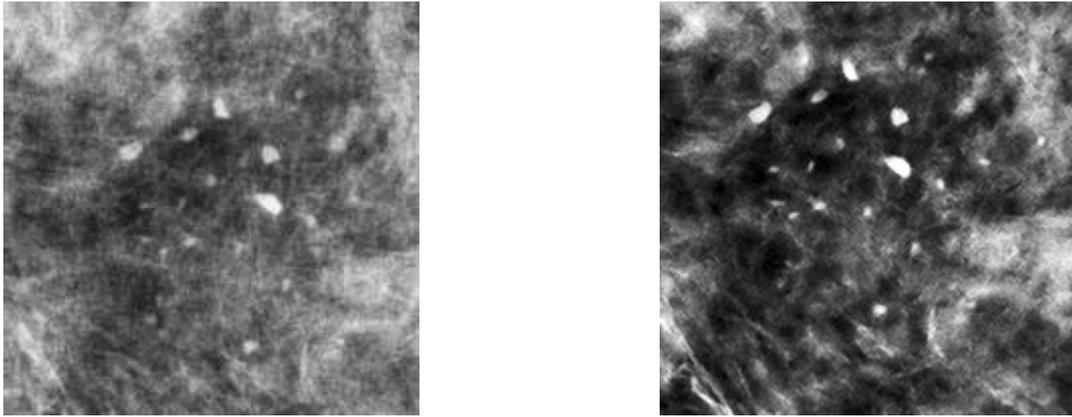
Figure 2. Section of TORMAM image acquired on same system five months apart following a change in image processing factors.

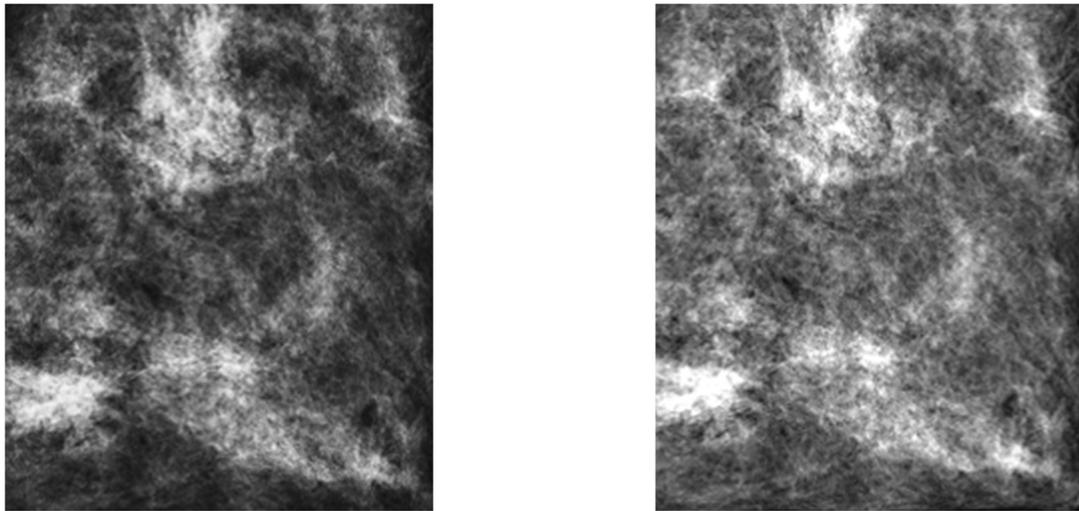
Figure 3. Section of TORMAM phantom for same model of equipment with different image processing set up differently between two centres.

**Dose levels**

The magnitude of the x-ray signal reaching the detector will be controlled by the automatic exposure control (AEC). There are different methods for selecting the cut-off point for an exposure and thereby the dose to the detector. The method will need to account for the radiographic factors selected, breast thickness and dense regions within the breast. Generally, there will be a default setting for a brand and type of system, however, some systems will have options to adjust the dose to a lower or higher value. In some systems this can be selected manually for an exposure and others it is set by the service engineer.

The GE Essential and Pristina have three dose levels, one of which is set as the default. Once the default level is set other dose levels can still be selected manually. In this survey, none of the systems used the low dose. Most used the standard (STD) dose level and a small number used the higher dose ('Contrast' for Essential, 'STD+' for Pristina).

The dose level for the AEC on Siemens systems can be configured by a service engineer. The standard is 100% and this value is chosen to obtain a specific image quality, i.e. 100% dose may be different between systems. The percentage dose level set for a system can be seen in tag 0018x,7062x and the results are summarised in Table 3. It should be noted that for the cases with reduced dose (<100%), these were all acquired using PRIME without a grid, which would be set up to be acquired at a lower dose. 32% of the systems were set up at the default dose of 100%.

Table 3. Number of dose levels for Siemens (number in bracket are PRIME)

| Dose Level | Inspiration | Revelation |
|---|---|---|
| <100% | 4 (4) | 1 (1) |
| 100% | 12 (0) | 13 (0) |
| 110%-119% | 7 (0) | 3 (3) |
| 120%-129% | 14 (2) | 5 (0) |
| 130%-140% | 0 (0) | 11 (0) |
| 150% | 6 (0) | 0 (0) |

**Physical differences**

Hologic changed their style of anti-scatter grid from a cellular design (HTC)[6] to a more standard linear grid. All of the 3Dimensions have a linear grid and 78% Selenia Dimensions have an HTC grid.

There are different styles of compression paddles such as flat, curved, flex, and FitSweet. These compression paddles causes the breast to be different in thickness on different positions on the breast support table. These differences are corrected for in the processing and could lead to differences in the perception of (clinical) images. A wide range on paddles were used by the centres in these surveys, but it should be noted that this does not necessarily mean that they are the paddles that are used clinically and a clinical audit will be required.

## 4. DISCUSSION

We had a very good response to the collection of data on image processing. In our unique dataset of 497 systems from across the UK. The main aim of this work was to examine the set-up of the systems as seen in images acquired with clinical processing. We have found a range of image processing versions, image processing settings, dose levels and physical set ups. There are limitations on the ability to compare the effects of these differences on image quality because TORMAMs themselves have a wide variation in internal configurations and appearances and are not particularly breast like and so we can only partially predict the effect on clinical images. . However, visual inspection of some example images demonstrated that these processing differences were not only seen in the header, but were also apparent in the appearance of the TORMAM images.

We have shown that each x-ray model had a number of different software versions that are currently in use across the UK. We do not know what all of the differences in software versions will mean to the appearance of clinical images. It is likely that some of the different version numbers will not affect image processing but only affect other parts of the system. However, we do know that past software updates have included changes to: AEC calibration curves, display of implants, skin edge appearance, contrast levels, flat fielding method, metal artefact reduction, and the appearance of tissue near the skin edge.

As has been demonstrated by Vries et al[4] processing can affect the outcome of AI, even at a level where readers do not detect a difference. It is reasonable to assume that other factors impacting the visual appearance of images or statistical distributions within the image data may similarly impact AI reader performance. There may even be differences by screening unit due to population differences across the country or variations in radiographic techniques such as compression paddles used and magnitude of compression.

All of these differences in image appearance may affect the implementation and continued use of AI in mammography screening units. The appearance of images used for the initial training of AI by the supplier may not be the same as used in a screening unit, indeed it is quite possible that training was undertaken on images acquired on systems with software versions/image processing that are no longer in use clinically. In practice, the introduction of AI will require a local training set perhaps not to retrain the AI but to set operating points. The training of an AI product for systems may face some difficulties if the systems have a range of appearances. Indeed, in a system with a particularly wide range of display features, this may mean collecting and training the AI using 'For Processing' images, although this would require a change to workflow. It is therefore essential that the AI is tested at installation and monitored throughout the lifetime of the product.

Inevitably, changes will be made to image processing, and/or physical set up or an entirely new system will be introduced. Consideration will need to be given on how to deal with such changes and it may need departments to stop using AI, until it has been established that the AI can function satisfactorily. It will mean that there is a need for close co-operation with the manufacturer and/or supplier of equipment for introducing changes and a clear disclosure of any changes that will affect the image appearance. There is a need for methods on testing AI.

The main purpose of this work was to look at the variability in image presentation that may affect AI. However, the information from this work will also be of interest in general to medical physics departments as they have responsibility for ensuring quality in equipment. They should be aware of these differences and this work may help them give advice to the imaging centres. We already know that many set up factors can affect cancer detection such as image processing, dose level and detector technology.[7–9] Quality control is undertaken by medical physics and screening units to ensure the quality of the images. However, generally there is little testing undertaken routinely on image processing by Medical Physics departments. This work provides some background knowledge for physics departments on what factors may change and what may be seen in the DICOM header.

Clinical departments should be undertaking clinical audit of the images and ensuring that the image processing is adequate across all of the subset of clinical images for different compressed breast thicknesses, glandularities and other features such as implants. This is not a simple process, but must be considered a multidisciplinary process with not only radiologists but also radiographers and physicists, some advice on the process can be found in EUREF.[10] There may be some value in undertaking audits of the settings of clinical images, although there are difficulties in extracting this sort of information from PACS.[11]

In addition to auditing the appearance of images, the outcomes (whether using AI or not) needs to be undertaken. It is a concern, that if mammography systems are changed in a way that can affect the outcome of the results from the AI, then it may take time before that effect can be shown. AI is a black box and we do not know what will influence the outcome and some understanding of this and we need to develop some methods to be able get an idea what influences the outcome.

A number of limitations must be noted, we have collected data from a large number of groups and there may be variations in how TORMAM images are acquired. Most of the data in the tags will be the same irrespective of the acquisition method. However, dose level may be affected by manual selection of settings. We have looked at a few examples of TORMAM images to see how the factors listed in this work can affect the image presentation. It is realised that the image processing might not function correctly for this phantom, but the images will give the reader some indication of the differences in processed images. The systems in this study were a mixture of screening, assessment and symptomatic units and the system may be set up differently according to its use. Tags were not always complete in the header and we know from experience that even completed tags can contain errors as has been previously shown.[12,13]

## 5. CONCLUSIONS

We have successfully shown that there are differences in the set-up of mammography systems across the UK. These differences will affect the appearance of the images, both to human readers and AI programs. As the AI systems are sensitive to the appearance of the images, then they will need to be set up for each screening unit, as the training set will likely to be different from the centre. The introduction of changes to the system or their set up will also need to be managed to ensure that the AI continues to function satisfactorily.

The differences in image appearance seen in this work may affect the clinical decisions made by image readers. Audit is required to ensure that the quality of images is sufficient for each screening unit. There are responsibilities on AI suppliers, physics, mammographic equipment manufacturers, and breast screening units to manage the use of AI and ensure the outcomes of breast screening are not adversely affected by the set-up of equipment.

## ACKNOWLEDGEMENTS

This work would not have been possible without Medical physics departments sending the data and in some cases follow up questions and images.

This work was supported by Project 22HLT05 MAIBAI, which has received funding from the European Partnership on Metrology, co-financed from the European Union's Horizon Europe Research and Innovation Programme and by the Participating States. Funding for the UK partners has been provided by Innovate UK under the Horizon Europe Guarantee Extension.

## REFERENCES

[1] Lamb, L. R., Lehman, C. D., Gastounioti, A., Conant, E. F. and Bahl, M., "Artificial Intelligence (AI) for Screening Mammography, From the *AJR* Special Series on AI Applications," American Journal of Roentgenology **219**(3), 369–380 (2022).

[2] Zanca, F., Jacobs, J., Ongeval, C. V., Claus, F., Celis, V., Geniets, C., Provost, V., Pauwels, H., Marchal, G. and Bosmans, H., "Evaluation of clinical image processing algorithms used in digital mammography," Med Phys **36**(3), 765–775 (2009).

[3] Warren, L. M., Given-Wilson, R. M., Wallis, M. G., Cooke, J., Halling-Brown, M. D., Mackenzie, A., Chakraborty, D. P., Bosmans, H., Dance, D. R. and Young, K. C., "The effect of image processing on the detection of cancers in digital mammography," American Journal of Roentgenology **203**(2), 387–393 (2014).

[4] De Vries, C. F., Colosimo, S. J., Staff, R. T., Dymiter, J. A., Yearsley, J., Dinneen, D., Boyle, M., Harrison, D. J., Anderson, L. A., Lip, G., on behalf of the iCAIRD Radiology Collaboration, Black, C., Murray, A. D., Wilde, K., Blackwood, J. D., Butterly, C., Zurowski, J., Eilbeck, J. and McSkimming, C., "Impact of Different Mammography Systems on Artificial Intelligence Performance in Breast Cancer Screening," Radiology: Artificial Intelligence **5**(3), e220146 (2023).

[5] Riddle, W. R. and Pickens, D. R., "Extracting data from a DICOM file," Med Phys **32**(6 Part1), 1537–1541 (2005).

[6] Moore, C. S., Wood, T. J., Saunderson, J. R. and Beavis, A. W., "A comparison of physical image quality of two hologic digital mammography systems that utilise linear and 2D anti-scatter grids," Biomed. Phys. Eng. Express **9**(5), 055017 (2023).

[7] Chiarelli, A. M., Edwards, S. A., Prummel, M. V., Muradali, D., Majpruz, V., Done, S. J., Brown, P., Shumak, R. S. and Yaffe, M. J., "Digital compared with screen-film mammography: performance measures in concurrent cohorts within an organized breast screening program," Radiology **268**(3), 684–693 (2013).

[8] Warren, L. M., Mackenzie, A., Cooke, J., Given-Wilson, R. M., Wallis, M. G., Chakraborty, D. P., Dance, D. R., Bosmans, H. and Young, K. C., "Effect of image quality on calcification detection in digital mammography," Medical physics **39**(6), 3202–3213 (2012).

[9] Mackenzie, A., Warren, L. M., Wallis, M. G., Cooke, J., Given-Wilson, R. M., Dance, D. R., Chakraborty, D. P., Halling-Brown, M. D., Looney, P. T. and Young, K. C., "Breast cancer detection rates using four different types of mammography detectors," European Radiology **26**(3), 874–883 (2016).

[10] EUREF., "European guidelines for quality assurance in breast cancer screening and diagnosis," 4th Edition-supplements (2013).

[11] Mackenzie, A., Lewis, E. and Loveland, J., "Successes and challenges in extracting information from DICOM image databases for audit and research," BJR **96**(1151), 20230104 (2023).

[12] Gueld, M. O., Kohnen, M., Keysers, D., Schubert, H., Wein, B. B., Bredno, J. and Lehmann, T. M., "Quality of DICOM header information for image categorization," Medical Imaging 2002: PACS and Integrated Medical Information Systems: Design and Evaluation **4685**, 280–287, SPIE (2002).

[13] Santos, M., Bastião, L., Costa, C., Silva, A. and Rocha, N., "Clinical Data Mining in Small Hospital PACS: Contributions for Radiology Department Improvement," J Am Coll Radiol **Ch.16**, 47–65 (2013).